# DNP surface- and subsurface-enhanced NMR spectroscopy of quadrupolar nuclei

Hiroki Nagashima,[1*] Julien Trébosc,[2,3] Yoshihiro Kon,[1] Kazuhiko Sato,[1] Olivier Lafon,[2,4*]
Jean-Paul Amoureux[2,5,6*]

[1] Interdisciplinary Research Center for Catalytic Chemistry, National Institute of Advanced Industrial
  Science and Technology (AIST), 1-1-1 Higashi, Tsukuba, Ibaraki 305-8565, Japan.
[2] Univ. Lille, CNRS, Centrale Lille, ENSCL, Univ. Artois, UMR 8181, Unit of Catalysis and Chemistry
  of Solids, F-59000 Lille, France.
[3] Univ. Lille, CNRS-2638, Fédération Chevreul, F-59000 Lille, France.
[4] Institut Universitaire de France, 1 rue Descartes, F-75231 Paris, France.
[5] Bruker Biospin, 34 rue de l'industrie, F-67166 Wissembourg, France.
[6] Riken NMR Science and Development Division, Yokohama, 230-0045 Kanagawa, Japan.

* Corresponding authors: nagashima-hiroki@aist.go.jp  olivier.lafon@univ-lille.fr  jean-
paul.amoureux@univ-lille.fr

**Abstract:** We introduce a novel approach, which extends the capabilities of dynamic nuclear polarization (DNP) to probe the local environment of half-integer spin quadrupolar nuclei. The use of this technique employing adiabatic dipolar recoupling instead of PRESTO or direct DNP greatly improves the sensitivity of DNP-NMR for the detection of quadrupolar isotopes with small dipolar couplings to protons, including notably those located in the subsurface of inorganic materials or with low gyromagnetic ratio ($\gamma$), such as $^{47,49}Ti$ or $^{95}Mo$. This technique was applied to probe, for the first time, the local environment of protonated and unprotonated $^{17}O$ sites in titania-supported $MoO_3$ ($MoO_3/TiO_2$), a heterogeneous catalyst widely used for the oxidation of hydrocarbons and alcohols as well as a precursor of hydro-desulfuration (HDS) catalysts. A high concentration of polarizing agent was employed because $MoO_3/TiO_2$ partially degrades it. The $^{17}O$ NMR spectra, acquired in natural abundance, indicate the presence of an uncoated $TiO_2$ surface associated to the formation of polyoxometalates or multi-layered $MoO_3$, as well as the existence of $HOMo_2$ and $HOMo_3$ acid sites. The enhanced sensitivity has also enabled the detection of $^{47,49}Ti$ or $^{95}Mo$ signals of $MoO_3/TiO_2$ and hence, the acquisition of the first DNP-enhanced NMR spectra of those low-$\gamma$ isotopes. These $^{47,49}Ti$ or $^{95}Mo$ NMR spectra show a partial disorder of $TiO_2$ and $MoO_3$ structures near the surface of the particles.

## 1. Introduction

  The surface of inorganic materials plays a key role in numerous applications, such as catalysis, biomaterials, cements and optoelectronics. Therefore, the determination of the structure-properties relationships for these surfaces is a powerful approach to design inorganic materials with improved properties. Nevertheless, this approach requires having appropriate tools to probe the structure of these surfaces. As a local characterization technique capable of providing atomic resolution, solid-state NMR can afford unique insights into the molecular-level structure of inorganic surfaces.[1–5] However, the sensitivity of solid-state NMR limits the characterization of surfaces, since only a small fraction of the nuclei are usually located near the surface.

Dynamic Nuclear Polarization (DNP), which consists in transferring the large polarization of unpaired electrons to surrounding nuclei, can boost the sensitivity of solid-state NMR.[6–10] This technique has notably been exploited to enhance the NMR signals of surfaces under Magic-Angle Spinning (MAS), first at $B_0 = 1.4$ T[11] and more recently at $B_0 \geq 9.4$ T in order to improve the spectral resolution.[12] This approach dubbed surface-enhanced NMR spectroscopy (SENS) has revealed novel surface structural features for a wide range of inorganic and hybrid materials, including meso- or micro-porous materials as well as nanoparticles, nanosheets and microparticles.[10,12–16]

The SENS strategy has often employed cross-polarization under MAS (CPMAS) sequences to transfer the DNP-enhanced $^1H$ polarization to other surface isotopes, including spin-1/2 isotopes, such as $^{13}C,$[12] $^{29}Si,$[17] $^{15}N,$[18] and $^{119}Sn,$[19] as well as half-integer spin quadrupolar isotopes, such as $^{27}Al,$[20] or $^{43}Ca.$[21] However, for half-integer quadrupolar nuclei, which represent around 66% of NMR active isotopes, the CPMAS scheme lacks robustness and efficiency and it is preferable to use the PRESTO technique, for which no spin lock is applied to the quadrupolar nucleus (Fig.S1).[22,23] The indirect DNP techniques, such as CPMAS and PRESTO, employing a coherent transfer of the DNP-enhanced $^1H$ polarization to the detected isotope, have the advantage that the $^1H$-$^1H$ spin diffusion transports the DNP-enhanced $^1H$ magnetization, which accelerates its build-up.

Furthermore, for inorganic materials, for which the bulk region contains no or a limited amount of protons, these indirect approaches allow the selective observation of nuclei near surfaces.[12,24] For instance, protonated surface $^{17}O$ sites, such as hydroxyl groups, have been selectively probed using $^1H \rightarrow ^{17}O$ CPMAS and PRESTO spectra.[22,25,26] The indirect DNP is more challenging for nuclei exhibiting weak dipolar coupling with protons, such as the non-protonated sites, including the subsurface ones, or the isotopes with low gyromagnetic ratio (γ). Furthermore, for abundant detected isotopes, such as $^{27}Al$ with 100% natural abundance (*NA*), the dipolar truncation, which affects both CPMAS and PRESTO schemes,[27,28] attenuates the transfers beyond the first neighbors. Hence, the $^1H \rightarrow ^{27}Al$ CPMAS and PRESTO schemes selectively transfer the [1]H polarization to the [27]Al nuclei in the first atomic layer of γ-alumina nanoparticles.[29]

An alternative to enhance the signal of un-protonated and/or subsurface sites is direct DNP, which involves the direct transfer of polarization from unpaired electrons to the detected isotope.[24,26,30–32] For slow relaxing nuclei, such as [29]Si, [119]Sn and [113]Cd, spin diffusion among them can transport the DNP-enhanced polarization from the surface to the bulk region.[33] This approach has been exploited in both direct and indirect DNP. However, it is generally not applicable for quadrupolar nuclei, which often exhibit fast relaxation because of the time fluctuations of the large quadrupole interaction.

We introduce herein a novel technique for the indirect DNP of half-integer quadrupolar nuclei. This technique employing an adiabatic recoupling scheme benefits from a fast build-up and a high robustness to [1]H-[1]H dipolar couplings. These properties improve the efficiency of polarization transfers through small dipolar couplings between protons and quadrupolar nuclei and hence, the sensitivity of the indirect DNP of subsurface sites and low-γ nuclei, such as [95]Mo or [47,49]Ti. We demonstrate a gain of 5.5 in sensitivity using the proposed indirect DNP technique instead of PRESTO, corresponding to an acceleration larger than 30 in the acquisition time. For the observation of non-protonated [17]O sites, indirect DNP is shown to reduce the acquisition time by a factor of 45 with respect to direct DNP. The introduced indirect DNP technique is applied to observe the [17]O, [47,49]Ti and [95]Mo NMR signals of titania-supported $MoO_3$ ($MoO_3/TiO_2$) in natural abundance. This heterogeneous catalyst is widely used for the oxidation of hydrocarbons and alcohols and the etherification of alcohols,[34,35] as well as a precursor of hydrodesulfuration (HDS) catalysts used to remove sulfur from petroleum refined products.[36,37] These NMR results notably provide new insights into the structure of $MoO_3/TiO_2$.

## 2. Methods

**2-1.    Sample preparation**. γ-alumina and α-$MoO_3$ were purchased from Sigma-Aldrich and Wako Pure Chemical Industries Ltd, respectively. Titanium oxide ($TiO_2$), commercialized under the name of P25 was obtained from Evonik Industries. Its surface area, measured by nitrogen sorption isotherms and calculated using the Brunauer-Emmett-Teller (BET) method, was equal to 62.5 $m^2.g^{-1}$. Ammonium molybdate tetrahydrate, $(NH_4)_6Mo_7O_{24}(H_2O)_4$, was obtained from Wako Pure Chemical Industries, Ltd. Two grams of $TiO_2$ P25 were added to 1 ml of $(NH_4)_6Mo_7O_{24}(H_2O)_4$ (245.3 mg, 0.2 mmol) aqueous solution. The solid was dried at 110 °C for 24 h, before being calcined at 500 °C for 3 h under air. After calcination, 2.20 g of $MoO_3/TiO_2$ catalyst containing about 10 wt% of $MoO_3$ has been obtained. For DNP-NMR experiments, the investigated samples were impregnated with TEKPol solutions in TCE during 1 h at room temperature. The mass of powder, the volume of the TEKPol solution and its concentration are given in Table S1. The samples were packed into 3.2 mm sapphire rotors.

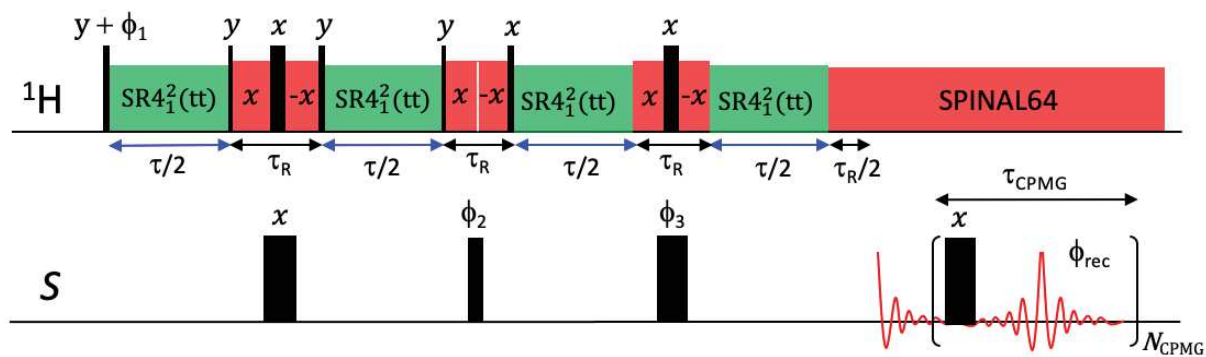

Figure 1. Pulse sequence of $^1H \rightarrow S$ RINEPT-SR4$_1^2$(tt)-QCPMG used to transfer the DNP-enhanced $^1$H polarization to the half-integer quadrupolar nucleus, S. The pulses on the S channel are selective of the central transition (CT). The narrow and wide black bars represent the $\pi/2$ and $\pi$ pulses, respectively. The total length of the recoupling time, the delay between successive $\pi$ pulses of QCPMG and the number of QCPMG echoes are denoted $\tau$, $\tau_{CPMG}$ and $N_{CPMG}$, respectively. The phase cycles are $\phi_1 = x$ -$x$, $\phi_2 = x\,x$ -$x$ -$x$, $\phi_3 = x\,x\,x\,x$ -$x$ -$x$ -$x$ -$x$ -$x\,y\,y\,y$ $y$ -$y$ -$y$ -$y$ -$y$, $\phi_{rec} = \phi_1 + \phi_2 - 2\phi_3$.

**2-2. MAS DNP-NMR experiments**. They were performed at 9.4 T on a Bruker BioSpin Avance NEO DNP-NMR spectrometer equipped with a double-resonance $^1$H/X 3.2 mm low-temperature MAS probe and a 263 GHz gyrotron.[38] The microwave irradiation was transmitted through a corrugated waveguide to the probe. The microwave power delivered to the sample was of ca. 6 W. The samples were spun at MAS frequencies of $\nu_R = 10$ or 12.5 kHz. The NMR spectra were acquired at a temperature of ca. 105 K, which was stabilized using a Bruker BioSpin MAS cooling system. The build-up curves of the $^1$H polarization with microwave irradiation were measured using a saturation-recovery experiment, involving a burst of 31 pre-saturation $\pi/2$-pulses separated by 1 ms, followed by a relaxation delay, $\tau_{RD}$, ranging from 1 to 7 s. After each relaxation delay, 1 transient was accumulated. The signal intensity of the saturation-recovery experiments was fitted as function of $\tau_{RD}$ to a single-exponential function. Such fit yielded the effective build-up time of the DNP-enhanced $^1$H polarization, $T_{B,on}(^1H)$, for the experiment with microwave irradiation. The sensitivity for the detection of the $^1$H signal was measured using saturation-recovery experiments with a relaxation delay $\tau_{RD} = 1.3T_{B,on}(^1H)$.[39]

We introduced a novel technique for the indirect DNP of half-integer quadrupolar nuclei, which is shown in Fig. 1. It is based on the dipolar-mediated refocused INEPT scheme (D-RINEPT), which is more robust to radio-frequency (rf) field inhomogeneity, offset and chemical shift anisotropy (CSA) of protons than PRESTO and is not affected by the dipolar truncation.[28] Furthermore, the robustness of this scheme, notably to $^1$H-$^1$H dipolar couplings, is enhanced by the use of an adiabatic dipolar recoupling,[40,41] based on the $SR4_1^2$ symmetry[42] and using the tanh/tan(tt) inversion pulse. As this $SR4_1^2$(tt) recoupling is non-γ-encoded, its blocks must be separated by an integer number of rotor periods in order to incorporate the $^1$H π/2 and π pulses. Continuous wave irradiations were applied during these delays in order to limit the losses due to the residual $^1$H-$^1$H dipolar couplings. The sensitivity can further be enhanced by the application of quadrupolar Carr-Purcell Meiboom-Gill (QCPMG) detection.[43]

We recorded $^1$H→$^{17}$O RINEPT-$SR4_1^2$(tt), RINEPT-$SR4_1^2$(tt)-QCPMG and PRESTO-R18$_2^5$-QCPMG spectra of γ-alumina impregnated with TEKPol solution, $^1$H→$^{95}$Mo RINEPT-$SR4_1^2$(tt)-QCPMG and PRESTO-R18$_2^5$-QCPMG spectra as well as $^1$H→$^{17}$O and $^1$H→$^{47,49}$Ti RINEPT-$SR4_1^2$(tt)-QCPMG spectra of MoO$_3$/TiO$_2$ impregnated with TEKPol solution. For these indirect DNP-NMR experiments, the $τ_{RD}$ delay was fixed to $1.3T_{B,on}(^1$H) in order to maximize the sensitivity.[39] The rf field of the pulses applied to the $^1$H channel, which do not belong neither to the recoupling scheme nor the CW irradiation, was equal to 137 kHz. The rf field of the CW irradiation was equal to 87 kHz. For the tanh/tan pulses of the $SR4_1^2$(tt) scheme, the rf field peak amplitude was 137 kHz, the peak amplitude of the carrier frequency modulation was 2 MHz, the ξ parameter, which determines the rise and the fall of the pulse, was 10, and the κ angle, which determines the slope of the frequency sweep in the middle of the pulse, was equal to 87°. The rf field of R18$_2^5$ recoupling was equal to 55 kHz. The rf field of the CT-selective π/2 and π pulses was equal to 13.9, 10.4 and 12.5 kHz for $^{17}$O, $^{95}$Mo and $^{47,49}$Ti isotopes. The direct excitation spectra under MAS of $^{17}$O, $^{95}$Mo and $^{47,49}$Ti nuclei in γ-alumina, α-MoO$_3$ and MoO$_3$/TiO$_2$ impregnated with TEKPol solutions were acquired using the DFS-QCPMG sequence.[43,44] The DFS pulse lasted 2 ms and used a rf nutation frequency of 35, 20 and 9 kHz for $^{17}$O, $^{95}$Mo and $^{47,49}$Ti nuclei, respectively. During the DFS pulse, the frequencies of the rf spikelets were linearly swept in a symmetric manner with respect to the CT from 800 to 250 kHz for $^{17}$O, 500 to 150 kHz for $^{95}$Mo, and 300 to 200 kHz for $^{47,49}$Ti. The rf field of the CT-selective π/2 and π pulses was identical to those used for indirect DNP experiments. For experiments with QCPMG detection, the spectra were obtained by co-adding the QCPMG echoes in the time domain, followed by a FT of the resultant echo. The 1D $^1$H→$^{13}$C CPMAS spectra of γ-alumina and MoO$_3$/TiO$_2$ impregnated with TEKPol solution were recorded using a $^1$H excitation π/2-pulse lasting 1.8 μs and a CP contact time of 1 ms. During the CP transfer, the rf nutation frequency on the $^{13}$C channel was constant and equal to 75 kHz, whereas that of $^1$H was linearly ramped from 70 to 100 kHz. The $^{17}$O isotropic chemical shifts were referenced to that of liquid $H_2O$ at 0 ppm. The [49]Ti isotropic chemical shifts were referenced to that of liquid neat $TiCl_4$ at 0 ppm using $SrTiO_3$ as secondary reference.[45] The [95]Mo isotropic chemical shifts were referenced to the [95]Mo resonance at 0 ppm of 2 mol.$L^{-1}$ aqueous solution of $Na_2MoO_4$. Additional experimental NMR parameters are given in Table S2.

**2-3.    EPR experiments.** They were performed using an X-band spectrometer operating at 9.62 GHz with respectively 100 kHz and 1.0 G for modulation frequency and amplitude. The microwave power was set to 0.316 mW to avoid saturation.

### 3.    Results and discussion

**3-1. Indirect DNP of unprotonated [17]O sites in γ-alumina**. Figure 2 shows the $^1H{\rightarrow}^{17}O$ RINEPT-SR$4_1^2$(tt) spectra of isotopically unmodified γ-alumina, i.e. with a [17]O abundance of 0.038%, impregnated with 5 mM TEKPol solution in 1,1,2,2-tetrachloroethane (TCE). For short τ delay, only the Al-[17]OH sites are observed, since the Al-[17]O-Al sites exhibit smaller $^1H$-[17]O dipolar couplings. Conversely, for τ = 1.12 ms the Al-[17]O-Al signal dominates (see Figure 2a). This result is the first observation of unprotonated oxygen nuclei using indirect DNP.[22,26,30] The low intensity of the Al-[17]OH sites for τ = 1.12 ms stems from (i) their smaller number than Al-[17]O-Al ones, since these are located both at the surface and in the subsurface region,[46] as well as (ii) the faster decay of the Al-[17]OH coherences because of stronger dipolar couplings with protons. The TEKPol concentration, $c_T$, was optimized to maximize the Al-[17]O-Al signal (see Figure S2). A low $c_T$ value, which lengthens the coherence lifetime, is favorable for the detection of unprotonated [17]O nuclei. As seen in Figure 2b, the RINEPT-SR$4_1^2$(tt)-QCPMG experiment allows the detection of Al-[17]O-Al signal with a sensitivity 3.9 times higher than that of PRESTO-R18$_2^5$-QCPMG, i.e. an acceleration of a factor of 15 in acquisition times. This higher sensitivity of RINEPT transfers, results from their better robustness, notably to $^1H$-$^1H$ dipolar couplings, because of the higher rf field used during SR$4_1^2$(tt) recoupling and CW irradiation than during R18$_2^5$ recoupling. For that sample, the RINEPT-SR$4_1^2$(tt)-QCPMG experiment yields a sensitivity for the detection of unprotonated [17]O nuclei 6.7 times higher than with direct DNP and double-frequency sweep (DFS) to enhance the polarization of the CT; i.e. an acceleration of a factor of 45 in acquisition times.

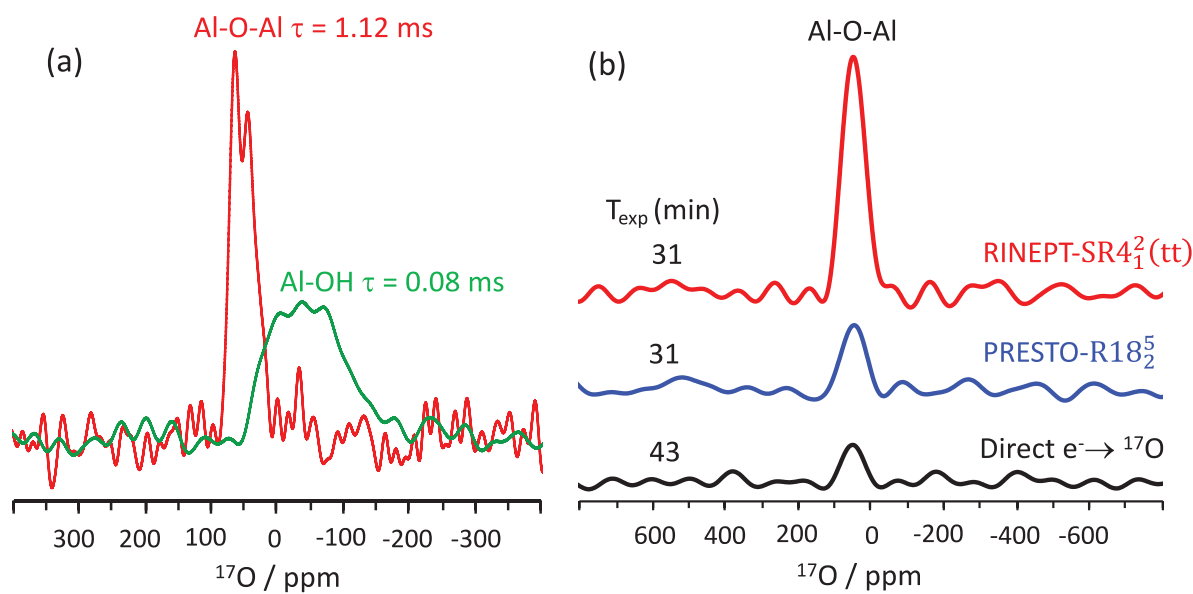

Figure 2. DNP-enhanced $^{17}O$ MAS spectra of isotopically unmodified γ-alumina impregnated with 5 mM TEKPol solution in TCE at $B_0$ = 9.4 T with MAS frequency $\nu_R$ = 12.5 kHz. (a) RINEPT-SR4$_1^2$($tt$) spectra with τ = 80 μs (green) and 1.120 ms (red) acquired in 5 h. (b) Comparison of $^{1}H{\rightarrow}^{17}O$ RINEPT-SR4$_1^2$($tt$)-QCPMG (top) and PRESTO-$R$18$_2^5$-QCPMG (middle) spectra with τ = 1.120 and 0.96 ms, respectively, with $^{17}O$ DFS-QCPMG spectra. The acquisition times of RINEPT-SR4$_1^2$($tt$)-QCPMG, PRESTO-$R$18$_2^5$-QCPMG and DFS-QCPMG spectra were equal to $T_{exp}$ = 31, 31 and 43 min, respectively. The spectra in (b) are the Fourier transform (FT) of the sum of QCPMG echoes.

## 3-2. Optimization of TEKPol concentration for MoO$_3$/TiO$_2$.

We then acquired the DNP-enhanced $^{1}H{\rightarrow}^{95}Mo$, $^{1}H{\rightarrow}^{17}O$ and $^{1}H{\rightarrow}^{47,49}Ti$ RINEPT-SR4$_1^2$(tt)-QCPMG spectra of isotopically unmodified MoO$_3$/TiO$_2$ impregnated with TEKPol solution in TCE. The concentration of TEKPol was optimized. 55 mM yielded the highest DNP enhancement for the $^{13}C$ signal of TCE in a $^{1}H{\rightarrow}^{13}C$ CPMAS experiment (see Table S3) and the highest sensitivity for the DNP-enhanced $^{1}H$ saturation-recovery experiment (see Figure S3), which is dominated by the $^{1}H$ signal of TCE. Conversely, a 110 mM TEKPol concentration provided a twice higher sensitivity, i.e. an acceleration of a factor of 4 in acquisition time, than 55 mM for $^{1}H{\rightarrow}^{95}Mo$ RINEPT-SR4$_1^2$(tt)-QCPMG experiments (see Figure S4). Such discrepancy between the optimal concentrations for the detection of TCE and $^{95}Mo$ nuclei could stem from a lower local concentration of TEKPol near the surface of MoO$_3$/TiO$_2$ particles than in the bulk of frozen TCE. Furthermore, the low sensitivity and DNP enhancement for TEKPol concentration of 27.5 mM (see Table S3 and Figure S3) suggest that a significant fraction of the TEKPol radical reacts with the MoO$_3$/TiO$_2$ surface.

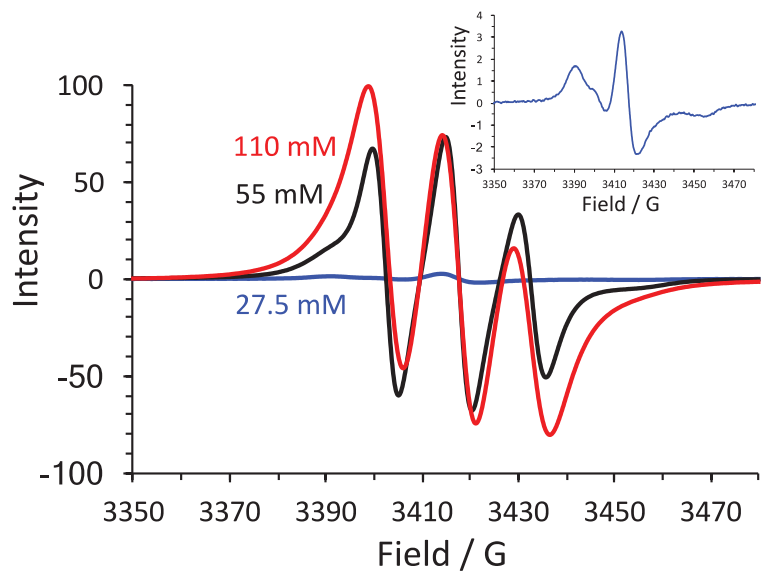

Figure 3 X-band EPR spectra of MoO$_3$/TiO$_2$ impregnated with 27.5 (blue), 55 (black) and 110 mM (red) TEKPol solutions in TCE at 298 K after an impregnation time of 1 h. The EPR spectrum of MoO$_3$/TiO$_2$ impregnated with 27.5 mM TEKPol solution is displayed at top right of the figure.

The decomposition of the TEKPol radicals was confirmed by the EPR measurements of unpaired electrons concentrations (see Table S3 and Figure 3), which is almost 10-fold lower for the sample impregnated with 27.5 mM solution than expected from the samples impregnated with 55 and 110 mM solutions. Besides, after impregnation with TEKPol solution, the DNP-enhanced $^1$H→$^{13}$C CPMAS spectra of MoO$_3$/TiO$_2$ samples exhibit additional $^{13}$C signals with respect to those of γ-alumina (see

Figure 4). The $^{13}$C isotropic chemical shifts of these signals correspond to those of diamagnetic parent molecules of TEKPol.[47] For instance, the nitroxide functions of the TEKPol radical could be reduced into hydroxylamines by reaction with the Mo(V) paramagnetic centers of MoO$_3$/TiO$_2$.[48] The increased intensity of the peak at 63 ppm for the 110 mM solution with respect to the 25 mM one may result from the hydroxylation of the methylene groups.

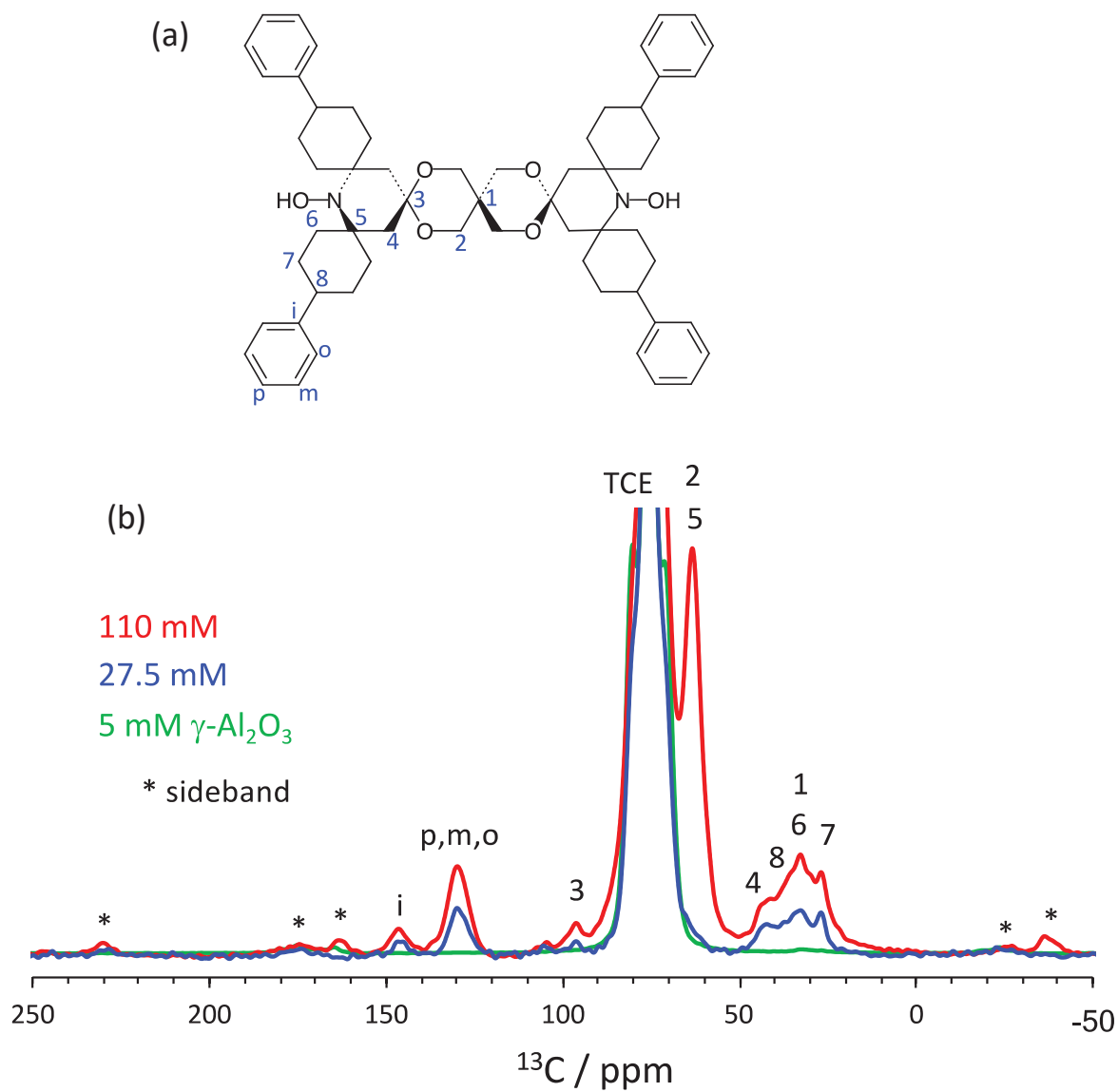

Figure 4 (a) Structure of hydroxylamine compound obtained by the reduction of TEKPol with atom numbering. (b) DNP-enhanced $^1$H→$^{13}$C CPMAS of of isotopically unmodified γ-alumina impregnated with 5 mM TEKPol solution in TCE (green) and isotopically unmodified MoO$_3$/TiO$_2$ impregnated with 27.5 (blue) and 110 mM (red) TEKPol solutions in TCE at $B_0$ = 9.4 T with $\nu_R$ = 12.5 kHz.

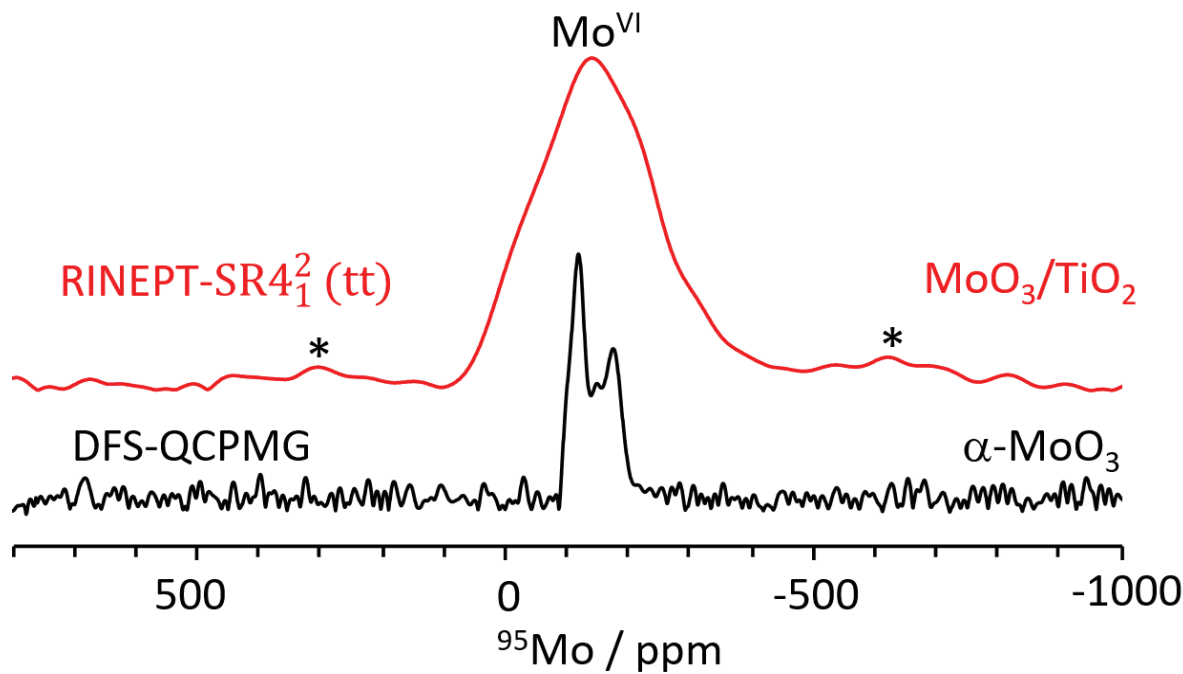

Figure 5. (top) DNP-enhanced $^1$H→$^{95}$Mo RINEPT-SR4$_1^2$($tt$)-QCPMG NMR spectrum of isotopically unmodified MoO$_3$/TiO$_2$ impregnated with 110 mM TEKPol solution in TCE with $\tau$ = 2.08 ms and (bottom) $^{95}$Mo DFS-QCPMG NMR spectrum of α-MoO$_3$ impregnated with 5 mM TEKPol solution in TCE at $B_0$ = 9.4 T with $\nu_R$ = 12.5 kHz. The spectra are the FT of the sum of QCPMG echoes.

### 3-3. $^{95}$Mo indirect DNP of MoO$_3$/TiO$_2$.

Figure 5**Erreur ! Source du renvoi introuvable.** shows the DNP-enhanced $^1$H→$^{95}$Mo RINEPT-SR4$_1^2$($tt$)-QCPMG NMR spectra of isotopically unmodified MoO$_3$/TiO$_2$ impregnated with 110 mM TEKPol solution in TCE, which is the first example of DNP-enhanced $^{95}$Mo spectrum reported in the literature.[10] As seen in Figure S5, the RINEPT-SR4$_1^2$($tt$) sequence is 5.5 times more sensitive than PRESTO-$R$18$_2^5$ for that sample, i.e. an acceleration of 30 in acquisition times. The higher sensitivity of RINEPT-SR4$_1^2$($tt$) stems notably from its shorter optimal recoupling time compared to PRESTO-$R$18$_2^5$.[28] The $^1$H→$^{95}$Mo RINEPT spectrum of MoO$_3$/TiO$_2$ exhibits a broad peak centered at 140 ppm, a shift similar to that of $^{95}$Mo DFS-QCPMG signal of α-MoO$_3$. This observation indicates that the predominant Mo sites in MoO$_3$/TiO$_2$ are hexa-coordinated in agreement with previous conventional solid-state NMR and XANES studies.[36,37,49] The broader $^{95}$Mo signal for MoO$_3$/TiO$_2$ with respect to α-MoO$_3$ results from (i) larger anisotropic interactions, such as the quadrupole and chemical shift anisotropy (CSA) ones, and (ii) a distribution of local environments.

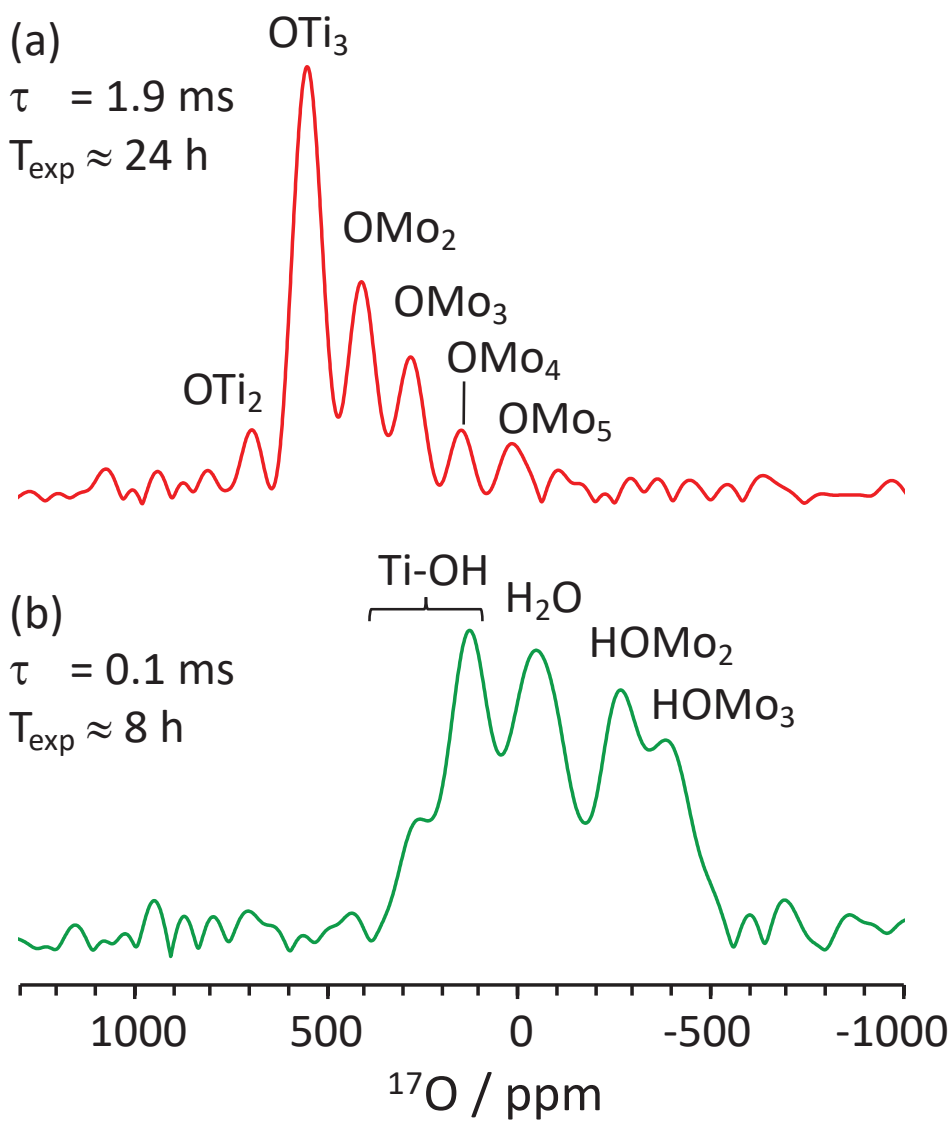

Figure 6. DNP-enhanced $^1H{\rightarrow}^{17}O$ RINEPT-$SR4_1^2$(tt)-QCPMG NMR spectra of isotopically unmodified $MoO_3/TiO_2$ impregnated with 110 mM TEKPol solution in TCE at $B_0 = 9.4$ T and $\nu_R = 10$ kHz with $\tau = $ (a) 1.9 and (b) 0.1 ms. The spectra are the FT of the sum of QCPMG echoes.

**3-4. $^{17}O$ indirect DNP of $MoO_3/TiO_2$.** RINEPT-SR4$_1^2$($tt$)-QCPMG sequence and 110 mM TEKPol concentration were also used to detect $^{17}O$ sites of $MoO_3/TiO_2$ (see Figure 6). To the best of our knowledge, $^{17}O$ NMR spectra of this catalyst have never been reported so far. Similarly to γ-alumina, the $^1H{\rightarrow}^{17}O$ RINEPT-SR4$_1^2$($tt$)-QCPMG spectra were recorded for short and long recoupling times in order to detect protonated and unprotonated oxygen sites. The $^{17}O$ spectrum for $\tau = 1.9$ ms shown in Figure 6a is dominated by a signal at 551 ppm, which is assigned to OTi$_3$ sites in the bulk region of anatase and rutile as well as those near the surface of TiO$_2$ particles.[50–53] The peak at 695 ppm is ascribed to OTi$_2$ sites at the surface of TiO$_2$ particles.[51,53] The higher intensity of the peak at 551 ppm with respect to that at 695 ppm demonstrates that RINEPT-SR4$_1^2(tt)$ allows probing subsurface O sites of $MoO_3/TiO_2$. The peaks at 420 and 285 ppm are assigned to $OMo_2$ and $OMo_3$ sites.[54] These sites are present in $[Mo_6O_{21}(Mo_4O_{12})_n]^{6-}$ clusters with $n \geq 2$, which have been identified by XANES spectroscopy as the predominant molybdenum oxide structures in $MoO_3/TiO_2$.[49] These signals can also be produced by $\alpha$-$MoO_3$ crystals formed at the surface of the $TiO_2$ particle.[55] The signals of OMoTi and $OMo_2Ti$ local environments are probably masked by those of $OTi_3$, $OMo_2$ and $OMo_3$ sites. The additional small peaks at 150 and 20 ppm can be ascribed to $OMo_4$ and $OMo_5$ sites, which may indicate the adsorption of polyoxometalates, such as $[Mo_7O_{24}]^{6-}$ and $\beta$-$[Mo_8O_{26}]^{4-}$, onto the particle surfaces.[54] The absence of terminal $^{17}O{=}Mo$ signals in the interval 900-750 ppm stems from their broadening by large CSA and quadrupole interactions.[5]

The $^1H{\rightarrow}^{17}O$ RINEPT-SR4$_1^2$(tt)-QCPMG experiment with $\tau = 0.1$ ms allows observing the protonated $^{17}O$ sites. The peaks at 130 and $-45$ ppm are assigned to TiOH and $H_2O$, respectively.[51,53] The signals at $-266$ and $-390$ ppm are ascribed to $HOMo_2$ and $HOMo_3$ sites, respectively.[56] The observation of TiOH signals indicates that the $TiO_2$ surface is not fully covered by 10 wt% molybdenum oxide, even if 6 wt% of $MoO_3$ should be sufficient to cover the surface of P25 particles by a monolayer of molybdenum oxide.[57] Hence, the presence of uncoated $TiO_2$ surface can be explained by the presence of polyoxometalates as well as the formation of multiple layers of $MoO_3$.

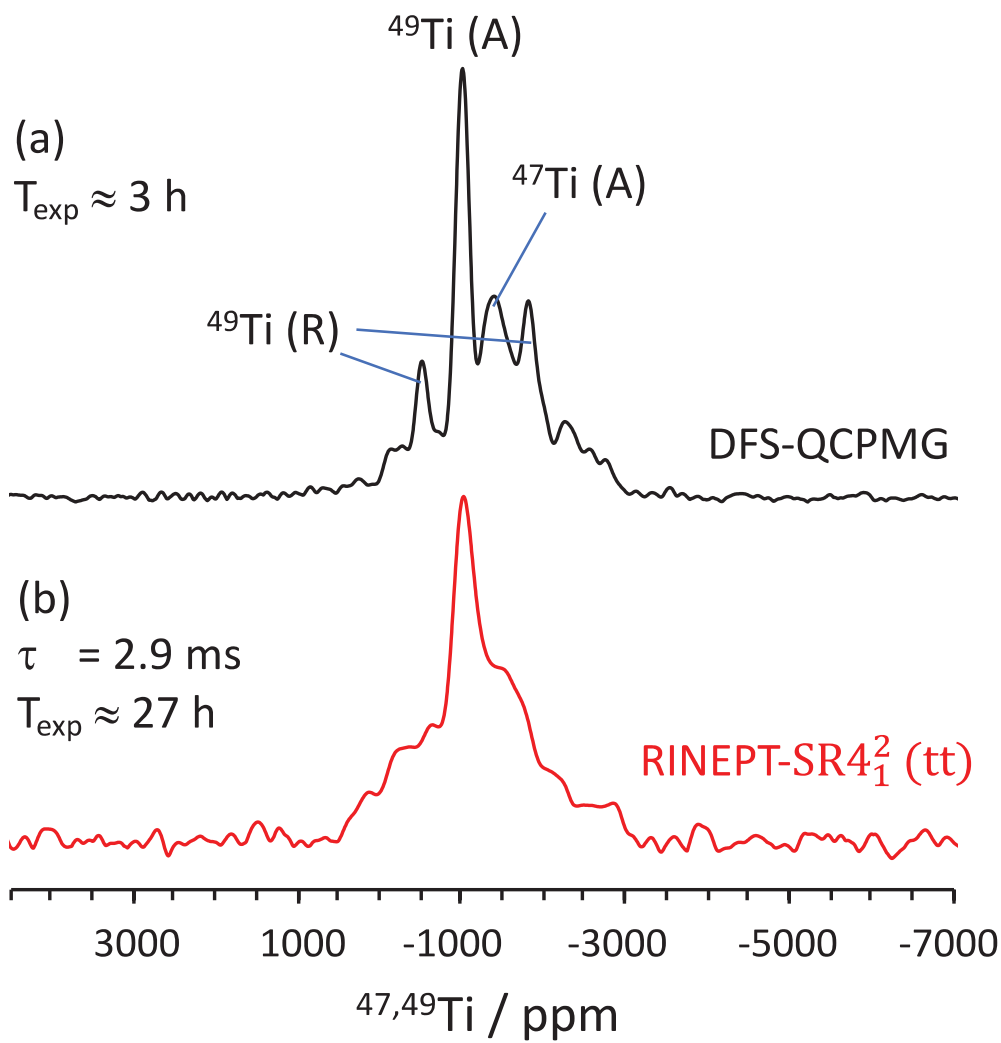

Figure 7. $^{47,49}$Ti NMR spectra of isotopically unmodified MoO$_3$/TiO$_2$ impregnated with 110 mM TEKPol solution in TCE at $B_0$ = 9.4 T with $\nu_R$ = 10 kHz acquired using (a) DFS-QCPMG and (b) $^1$H→$^{47,49}$Ti RINEPT-SR4$_1^2$(tt)-QCPMG with $\tau$ = 2.9 ms.

**3-5. $^{47,49}$Ti indirect DNP of MoO$_3$/TiO$_2$.** The $^{47,49}$Ti DFS-QCPMG spectrum of isotopically unmodified MoO$_3$/TiO$_2$ is shown in Figure 7a. It is dominated by the signal of $^{49}$Ti nuclei in the bulk anatase ($^{49}$Ti(A)).[45,58] NMR signals of $^{47}$Ti nuclei in the bulk anatase ($^{47}$Ti(A)) and $^{49}$Ti nuclei in the bulk rutile ($^{49}$Ti(R)) are also observed. The signal of $^{47}$Ti(R) is too broad to be resolved. Figure 7b shows the DNP-enhanced $^1$H→$^{47,49}$Ti RINEPT-SR4$_1^2$(tt)-QCPMG spectrum of the same sample. It is the first example of $^{47,49}$Ti DNP-NMR spectrum reported in the literature.[10] $^1$H→$^{47,49}$Ti transfer allows selecting the $^{47,49}$Ti nuclei near the surface. However, given the low gyromagnetic ratios ($\gamma(^{47}$Ti$)/\gamma(^1$H$)$ = 0.05638 and $\gamma(^{49}$Ti$)/\gamma(^1$H$)$ = 0.05639), the low natural abundance ($NA(^{47}$Ti$)$ = 5.41% and $NA(^{49}$Ti$)$ = 7.44%) of these isotopes, and the limited surface area of P25 (62.5 m².g⁻¹), the DNP sensitivity gain is required to detect the NMR signal of the $^{47,49}$Ti nuclei near the surface. The most intense peak in this spectrum resonates at the same shift as $^{49}$Ti(A). Nevertheless, the transitions detected in the RINEPT spectrum are significantly broader than in Figure 7a. Such broadening results from larger quadrupole interactions and a distribution in the local environment for nuclei near surfaces with respect to those in the bulk region.

### 4. Conclusions

In this article, we have introduced the RINEPT-SR4$_1^2$(tt) technique and demonstrated that it can be used to enhance the DNP signals of quadrupolar nuclei, such as $^{17}$O, for species both at the surface and subsurface. We have shown that this technique is more efficient than PRESTO and direct DNP for the observation of unprotonated $^{17}$O sites. This method has been applied to record the first $^{17}$O NMR spectra of $MoO_3/TiO_2$. A high TEKPol concentration was used to compensate the partial degradation of this polarizing agent by a reaction with the $MoO_3/TiO_2$. The obtained $^{17}$O spectra indicate that (i) $TiO_2$ particles are not fully coated with $MoO_3$, (ii) molybdenum oxide forms multiple layers or polyoxometalates, and (iii) $MoO_3/TiO_2$ catalyst contains $HOMo_2$ and $HOMo_3$ acid sites. The fast build-up and the robustness to $^1$H-$^1$H dipolar couplings of RINEPT-SR4$_1^2$(tt) method is also advantageous for the observation of low-$\gamma$ quadrupolar nuclei and we have applied this technique to record the first DNP-NMR spectra of $^{95}$Mo and $^{47,49}$Ti. We have notably shown that RINEPT-SR4$_1^2$(tt) accelerates the acquisition of $^{95}$Mo NMR spectrum of $MoO_3/TiO_2$ by a factor of 30 with respect PRESTO-R18$_2^5$. The DNP-enhanced $^1$H$\rightarrow$$^{47,49}$Ti RINEPT-SR4$_1^2$(tt) spectrum also indicates a partial disorder of anatase and rutile structures near the surface.

**Acknowledgments**

HN thank the New Energy and Industrial Technology Development Organization (NEDO) for the support of this research (Grant No. P16010). Chevreul Institute (FR 2638), Ministère de l'Enseignement Supérieur, de la Recherche et de l'Innovation, Hauts-de-France Region and FEDER are acknowledged for supporting and funding partially this work. Financial support from the IR-RMN-THC FR-3050 CNRS for conducting the research is gratefully acknowledged. Authors also thank contracts ANR-17-ERC2-0022 (EOS) and ANR-18-CE08-0015-01 (ThinGlass). This project has received funding from the European Union's Horizon 2020 research and innovation program under grant agreement No 731019 (EUSMI). OL acknowledge financial support from Institut Universitaire de France (IUF). HN would like to thank Takuya Nakashima for the preparation of 10 wt% $MoO_3$-$TiO_2$ catalyst.

# Supporting Information

# DNP surface- and subsurface-enhanced NMR spectroscopy of quadrupolar nuclei

Hiroki Nagashima,[1*] Julien Trébosc,[2,3] Yoshihiro Kon,[1] Kazuhiko Sato,[1] Olivier Lafon,[2,4*]
Jean-Paul Amoureux[2,5,6*]

[1] Interdisciplinary Research Center for Catalytic Chemistry, National Institute of Advanced Industrial
Science and Technology (AIST), 1-1-1 Higashi, Tsukuba, Ibaraki 305-8565, Japan.
[2] Univ. Lille, CNRS, Centrale Lille, ENSCL, Univ. Artois, UMR 8181, Unité de Catalyse et de Chimie
du Solide, F-59000 Lille, France.
[3] Univ. Lille, CNRS-2638, Fédération Chevreul, F-59000 Lille, France.
[4] Institut Universitaire de France, 1 rue Descartes, F-75231 Paris, France.
[5] Bruker Biospin, 34 rue de l'industrie, F-67166 Wissembourg, France.
[6] Riken NMR Science and Development Division, Yokohama, 230-0045 Kanagawa, Japan.

* Corresponding authors: nagashima-hiroki@aist.go.jp  olivier.lafon@univ-lille.fr, jean-
paul.amoureux@univ-lille.fr

## Experimental section

Table S1. Composition of the samples investigated by NMR, including the mass of the powder ($m$), the volume
of the TEKPol solution ($V$) and its concentration ($c$(TEKPol)).

| Sample | $m$ /mg | $V$ /μL | $c$(TEKPol) /mM |
|---|---|---|---|
| $\gamma$-Al$_2$O$_3$ | 21 | 30 | 5, 10 or 16 |
| MoO$_3$/TiO$_2$ | 40 | 20 | 5, 16, 27.5, 55 or 110 |
| $\alpha$-MoO$_3$ | 80 | 20 | 5 |

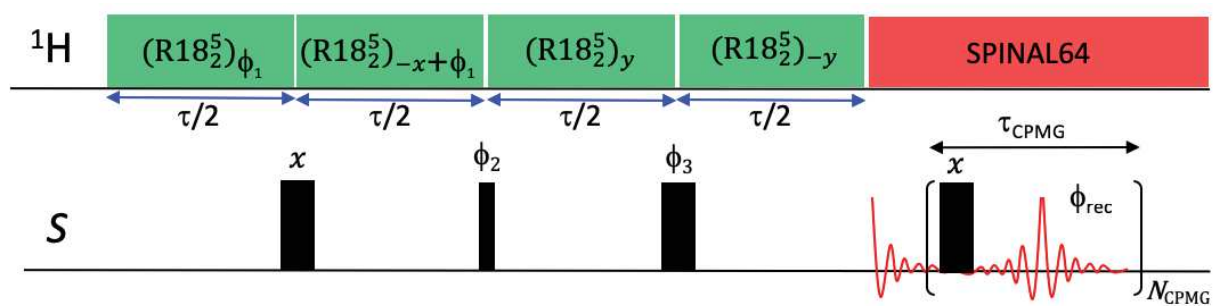

Figure S1. Pulse sequence of PRESTO-$R18_2^5$-QCPMG used to transfer the DNP-enhanced $^1$H polarization to half-
integer quadrupolar nucleus $S$. The pulses on the $S$ channel are selective of the central transition (CT). The narrow
and wide black bars represent the π/2 and π pulses, respectively. The phase cycles are $\phi_1 = x$ -$x$, $\phi_2 = x$ $x$ -$x$ -$x$, $\phi_3 = x$
$x$ $x$ $x$ -$x$ -$x$ -$x$ -$x$ $y$ $y$ $y$ $y$ -$y$ -$y$ -$y$ -$y$, $\phi_{rec} = \phi_1 + \phi_2 - 2\phi_3$.

Table S2. Experimental parameters used for the acquisition of the NMR spectra shown in this article. τ: total length
of recoupling time, $\nu_R$: MAS frequency, $T_{CPMG}$: delay in between two pulses of the QCPMG train, $N_{CPMG}$: number of echoes, $C_T$: TEKPol concentration in TCE, $\varepsilon_{on/off\,(1H)}$: DNP enhancement, $\tau_{RD}$: recycle delay, NS: number of scan, $T_{exp}$: acquisition time.

| Fig | Pulse sequence | $\tau$ /ms | $\nu_R$ /kHz | $T_{CPMG}$ /ms | $N_{CPMG}$ | $C_T$ /mM | $\varepsilon_{on/off}$ ($^1$H) [a] | $\tau_{RD}$ /s | $NS$ | $T_{exp}$ |
|---|---|---|---|---|---|---|---|---|---|---|
| **2a** | RINEPT-SR4$_1^2$(tt) | 0.08 | 12.5 | - | - | 5 | 125 | 7.2 | 3072 | 4 h |
| | | 1.12 | 12.5 | - | - | 5 | 125 | 7.2 | 1984 | 6.2 h |
| **2b** | RINEPT-SR4$_1^2$(tt)-QCPMG | 1.12 | 12.5 | 0.48 | 100 | 5 | 125 | 7.2 | 256 | 31 min |
| | PRESTO-R18$_2^5$-QCPMG | 1.12 | 12.5 | 0.48 | 100 | 5 | 125 | 7.2 | 256 | 31 min |
| | DFS-QCPMG (Direct DNP) | - | 12.5 | 0.48 | 100 | 5 | - | 7.2 | 256 | 43 min |
| **3** | RINEPT-SR4$_1^2$(tt)-QCPMG | 2.08 | 12.5 | 1.12 | 43 | 110 | 76 | 2 | 6144 | 3.4 h |
| | DFS-QCPMG | - | 12.5 | 3.84 | 30 | 5 | - | 15 | 180 | 45 min |
| **4a** | RINEPT-SR4$_1^2$(tt)-QCPMG | 1.9 | 10 | 0.4 | 122 | 110 | 76 | 1 | 86400 | 24 h |
| **4b** | RINEPT-SR4$_1^2$(tt)-QCPMG | 0.1 | 10 | 0.4 | 25 | 110 | 76 | 1 | 28800 | 8 h |
| **5a** | DFS-QCPMG | - | 10 | 0.8 | 248 | 110 | - | 1 | 9000 | 3 h |
| **5b** | RINEPT-SR4$_1^2$(tt)-QCPMG | 2.9 | 10 | 0.8 | 61 | 110 | 76 | 1 | 97200 | 27 h |
| **S2** | RINEPT-SR4$_1^2$(tt)-QCPMG | 0.08 | 12.5 | 0.48 | 20 | 5 | 125 | 7.2 | 256 | 31 min |
| | | 0.08 | 12.5 | 0.48 | 20 | 10 | 221 | 2.7 | 768 | 34 min |
| | | 0.08 | 12.5 | 0.48 | 20 | 16 | 256 | 2.6 | 512 | 23 min |
| | | 1.12 | 12.5 | 0.48 | 100 | 5 | 125 | 7.2 | 256 | 31 min |
| | | 0.96 | 12.5 | 0.48 | 100 | 10 | 221 | 2.7 | 768 | 34 min |
| | | 0.96 | 12.5 | 0.48 | 100 | 16 | 256 | 2.6 | 512 | 23 min |
| **S3** | RINEPT-SR4$_1^2$(tt)-QCPMG | 2.08 | 12.5 | 0.48 | 100 | 55, | 157 | 2 | 512 | 17 min |
| | | 2.08 | 12.5 | 0.48 | 100 | 110 | 76 | 2 | 512 | 17 min |
| **S5** | Ramped CP | 1 | 10 | - | - | 5 $\gamma$-Al$_2$O$_3$ | 125 | 7.2 | 64 | 8 min |
| | | 1 | 10 | - | - | 27.5 | 6.8 | 15.4 | 64 | 16 min |
| | | 1 | 10 | - | - | 110 | 76 | 1.5 | 512 | 13 min |
| **S6** | RINEPT-SR4$_1^2$(tt)-QCPMG | 2.08 | 12.5 | 0.48 | 102 | 110 | 76 | 2 | 256 | 17 min |
| | PRESTO-R18$_2^5$-QCPMG | 2.56 | 12.5 | 0.48 | 102 | 110 | 76 | 2 | 256 | 17 min |

[a] DNP enhancement of the $^1$H signals of TCE calculated as the ratio between the $^1$H signals of TCE in $^1$H Hahn echo spectra with and without microwave irradiation.

## Optimization of TEKPol concentration for γ-alumina

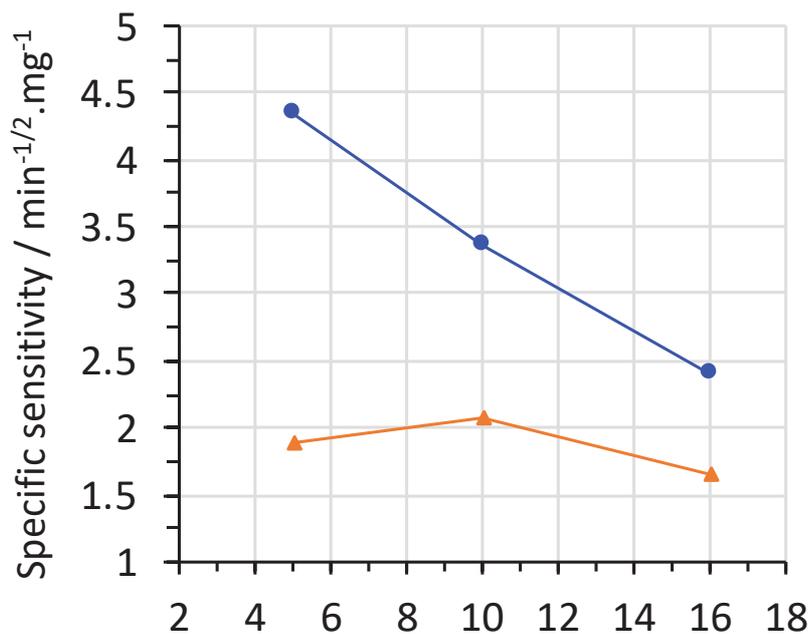

Figure S2. Relative specific sensitivity $\kappa$ in / min$^{-1/2}$.mg$^{-1}$, i.e. the signal-to-noise ratio divided by the mass of sample and the square root of the acquisition time, as function of TEKPol concentration for Al-$^{17}$O-Al (blue disk) and Al-$^{17}$OH (brown triangle) signals in $^{1}$H→$^{17}$O RINEPT-SR4$_1^2$(tt)-QCPMG of isotopically unmodified γ-alumina impregnated with TEKPol solution in TCE with τ = 80 μs for Al-$^{17}$OH and τ = 1.12 ms for Al-$^{17}$O-Al. The continuous line is only a guide for the eyes.

## Optimization of TEKPol concentration for MoO₃/TiO₂

Table S3. Mass concentration of unpaired electrons, $\rho_e$, $\rho_e/c_T$ and DNP enhancement, $\varepsilon_{on/off}$(TCE), for MoO₃/TiO₂ samples impregnated with 27.5, 55 and 110 mM TEKPol solution in TCE.

| c(TEKPol) /mM | 27.5 | 55 | 110 |
|---|---|---|---|
| $\rho_e$ /spin.g$^{-1}$ $^a$ | $1.0 \times 10^{18}$ | $2.6 \times 10^{19}$ | $3.8 \times 10^{19}$ |
| $\rho_e$/c(TEKPol) / spin.g$^{-1}$.mM$^{-1}$ | $3.8 \times 10^{16}$ | $4.5 \times 10^{17}$ | $3.5 \times 10^{17}$ |
| $\varepsilon_{on/off}$(TCE) $^b$ | 6 | 236 | 173 |

$^a$ Determined from the integrated intensity of the EPR signal. $^b$ Calculated as the ratio between the $^{13}$C signals of TCE in $^{1}$H→$^{13}$C CPMAS spectra with and without microwave irradiation.

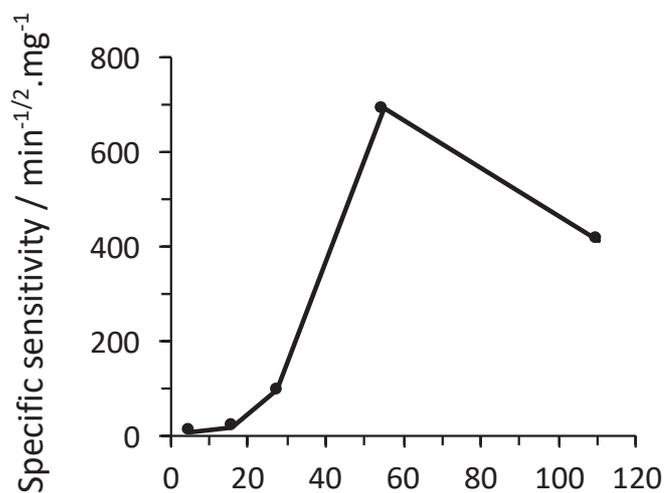

Figure S3. Relative specific sensitivity in min$^{-1/2}$.mg$^{-1}$ as function of TEKPol concentration for $^1$H signal of isotopically unmodified MoO$_3$/TiO$_2$ impregnated with TEKPol solution in TCE impregnated. The continuous line is only a guide for the eyes.

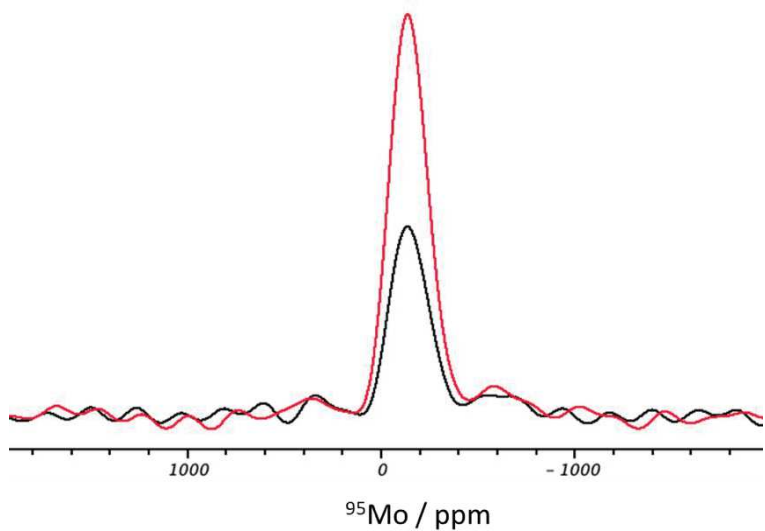

Figure S4. DNP-enhanced $^1$H→$^{95}$Mo RINEPT-SR4$_1^2$(tt)-QCPMG NMR spectra of isotopically unmodified MoO$_3$/TiO$_2$ impregnated with 55 (black) and 110 mM (red) TEKPol solutions in TCE at $B_0$ = 9.4 T with $\nu_R$ = 12.5 kHz.

**Comparison of $^1H \rightarrow ^{95}Mo$ PRESTO and RINEPT sensitivity**

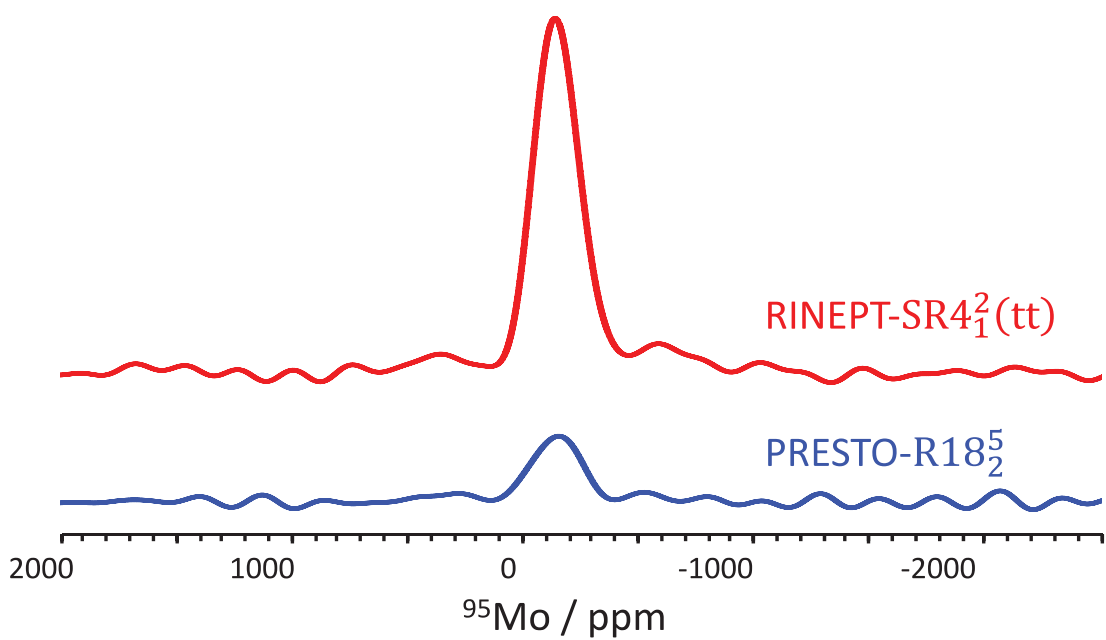

Figure S5. DNP-enhanced $^1H \rightarrow ^{95}Mo$ RINEPT-SR4$_1^2$($tt$)-QCPMG (top) and PRESTO-$R$18$_2^5$-QCPMG (bottom) NMR spectra of isotopically unmodified MoO$_3$/TiO$_2$ impregnated with 110 mM TEKPol solution in TCE. The τ delay is equal to 1.9 and 2.6 ms for RINEPT and PRESTO experiments, respectively. The spectra are the FT of the sum of QCPMG echoes. The $^1H \rightarrow ^{95}Mo$ RINEPT-SR4$_1^2$($tt$)-QCPMG spectrum is the same as that displayed in Figure 5.